\documentclass[twocolumn,floatfix,preprintnumbers,citeautoscript,newabstract,nofootinbib]{revtex4} 
\usepackage{amsmath,amssymb,graphics}
\usepackage{bbm}
\newcommand{\bs}[1]{\boldsymbol{#1}}
\newcommand{\bsb}[1]{\boldsymbol{\overline{#1}}}

\begin{document} \setlength{\unitlength}{1in}

\preprint{OHSTPY-HEP-T-07-001\\[0.5cm] }

\title{Gauge coupling unification and light Exotica in String Theory\\[1cm]}

\author{{\bf\normalsize
Stuart~Raby and
Ak{\i}n~Wingerter}\\[0.5cm]
{\it\normalsize
Department of Physics, The Ohio State University,}\\[-0.05cm]
{\it\normalsize
191 W.\ Woodruff Ave., Columbus, OH 43210, USA}\\[0.15cm]
}

\begin{abstract}
In this letter we consider the consequences for the LHC of light
vector-like exotica with fractional electric charge.   It is shown
that such states are found in orbifold constructions of the
heterotic string.   Moreover, these exotica are consistent with
gauge coupling unification at one loop,  even though they do not
come in complete multiplets of SU(5).
\end{abstract}

\pacs{\dots}

\maketitle

The LHC era is about to begin.   The minimal supersymmetric standard
model [MSSM] is the premier candidate for new physics beyond the
Standard Model [SM], providing a plethora of new signatures for LHC
physics.  Many attempts have been made to derive the MSSM from
string theory.    Such attempts typically lead to the MSSM spectrum
plus additional states, termed {\em exotics}.

D-brane constructions typically contain exotics which are {\em
chiral} under the Standard Model gauge group
\cite{Blumenhagen:2005mu}.   As a result, in the best of
circumstances these chiral exotics obtain mass of order the weak
scale, only after electroweak symmetry breaking.  In the worst of
circumstances, the chiral exotics remain massless. Hence these
theories with {\em chiral exotics} are severely constrained by
precision electroweak data \cite{chiralexotics}.

Recently,  heterotic string constructions using orbifold
compactifications have provided models which have only the MSSM
spectrum or the MSSM spectrum with the addition of {\em vector-like
exotics}
\cite{Kobayashi:2004ud,Forste:2004ie,Kobayashi:2004ya,Buchmuller:2005jr,Buchmuller:2006ik,Buchmuller:2004hv,Buchmuller:2005sh,Lebedev:2006kn}.
Vector-like exotics can by definition obtain mass at an arbitrary
scale without breaking any Standard Model gauge symmetry. Simple
examples of vector-like exotics include additional pairs of Higgs
doublets (${\bf H_u + H_d}$), a pair of states with the quantum
numbers of a down quark, ${\bf D^c}$, $(\bar 3, 1)_{\frac{2}{3}}$
and its charge conjugate or complete multiplets under $\text{SU}(5)$
such as $\bsb{5} + \bs{5}$, transforming under the SM as $(\bsb{3},
\bs{1})_{\frac{2}{3}} + (\bs{1}, \bs{2})_{-1}$ [${\bf D^c + L}$] +
c.c. In the first example,  these states contribute to the
renormalization group running of the SM gauge coupling in such a way
as to change the GUT scale.   If these states are light, with mass
of order the weak scale,  then they will destroy the nice agreement
of the LEP data and gauge coupling unification in SUSY GUTs. Hence,
in order to preserve the nice features of SUSY GUTs, one typically
argues that these states have mass of order the GUT scale. On the
other hand, if the vector-like exotics come in complete
$\text{SU}(5)$ multiplets, then at least at one loop they do not
affect gauge coupling unification, even if they are light.  In fact,
light vector-like exotics in complete $\text{SU}(5)$ multiplets have
been used as messengers for low energy gauge-mediated SUSY
breaking~\cite{gmsb,gmsb2,gmsb3,Giudice:1998bp}.

In this paper we point out that it is also possible to have light
vector-like exotics,  which are {\em not} in complete $\text{SU}(5)$
multiplets, however still preserving the nice features of SUSY GUTs
\cite{drw,raby}. Moreover, in the simple example, these states have
fractional electric charge and thus it would be very interesting to
search for such states at the LHC.  Such states are present in many
heterotic orbifold constructions.

\section{Stringy Exotica}

In  Ref.~\cite{Kobayashi:2004ud,Kobayashi:2004ya} the $\text{E}_8
\times \text{E}_8$ heterotic string was compactified on a particular
six torus modded out by the discrete symmetry $\mathbb{Z}_6$-II $=
\mathbb{Z}_3 \times \mathbb{Z}_2$.   By taking one compactified
dimension to be much larger than the other 5, it was shown in one of
the examples that the $\mathbb{Z}_3$ orbifold by itself produced a
low energy effective field theory (below the string scale)
equivalent to an $\text{E}_6$ orbifold GUT in 5 space-time
dimensions.   The final $\mathbb{Z}_2$ orbifolding of the one extra
dimension, along with one Wilson line, reduced the $\text{E}_6$
gauge symmetry to Pati-Salam [PS] $\text{SU}(4) \times
\text{SU}(2)_L \times \text{SU}(2)_R$. The resulting orbifold
described a line segment with two end-of-the-world branes.   At one
end the gauge symmetry was $\text{SO}(10)$, while at the other end
it was $\text{SU}(6) \times \text{SU}(2)_R$. Whereas Standard Model
families resided on the $\text{SO}(10)$ brane,  it was found that
vector-like exotics in the representations, $[(\bs{6}, \bs{1}) +
(\bs{1}, \bs{2})]$ + charge conj.,  were localized on the
$\text{SU}(6) \times \text{SU}(2)_R$ brane. Under PS, i.e.
$$\text{SU}(6) \times \text{SU}(2)_R \longrightarrow \text{SU}(4) \times \text{SU}(2)_L \times
\text{SU}(2)_R,$$
these states transform as
\begin{eqnarray} & [(\bs{6}, \bs{1}) + (\bs{1}, \bs{2})] +
\text{c.c.} & \\ & \longrightarrow  [(\bs{4}, \bs{1}, \bs{1}) + (\bs{1}, \bs{2}, \bs{1}) + (\bs{1}, \bs{1}, \bs{2})] +
\text{c.c.} & \nonumber \end{eqnarray}
Assuming that PS spontaneously breaks to
the Standard Model, i.e.
$$\text{SU}(4) \times \text{SU}(2)_L \times \text{SU}(2)_R \longrightarrow \text{SU}(3)_c
\times \text{SU}(2)_L \times \text{U}(1)_Y,$$
we finally obtain
\begin{eqnarray} & [(\bs{4}, \bs{1}, \bs{1}) + (\bs{1}, \bs{2}, \bs{1}) + (\bs{1}, \bs{1}, \bs{2})]  + \text{c.c.}    & \label{eq:exotica} \\
&\longrightarrow [(\bs{3}, \bs{1})_{1/3} +
(\bs{1}, \bs{1})_{-1} + (\bs{1}, \bs{2})_0 + (\bs{1}, \bs{1})_{\pm 1}] +  \text{c.c.}  &  \nonumber \\
& = \;\; [{\bf Q + E_- + L + E_\pm}] + \text{c.c.}  & \nonumber
\end{eqnarray}
with $Y = (B - L) + 2 T_{3 R}$ and electric charge $Q = T_{3 L} +
Y/2$.
Note, these states are explicitly {\em vector-like}.  They are all
located in the same twisted sector of the string.  Thus it is not
unreasonable to expect that they all obtain mass (when SM singlet
fields get supersymmetric vacuum expectation values) of the same
order.  We shall therefore explore the possibility that they have a
gauge invariant, supersymmetric mass ${\bf M}$ of order the
electroweak scale.\footnote{We assume here that all these exotics
obtain the same mass ${\bf M}$, however this is not a priori
necessary.} In addition, they are {\em very exotic}; henceforth we
refer to them as {\em exotica}.  They have fractional electric
charge given by
\begin{eqnarray}
&[{\bf Q_{1/6} + E_{-1/2} + L_{\pm 1/2} + E_{\pm 1/2}}] + \text{c.c.}  &
\end{eqnarray}
The exotic leptons have charges $\pm 1/2$,  while the bound states
of exotic quarks with SM quarks form an iso-vector and iso-scalar
multiplet of baryons
\begin{eqnarray} {\bf \Sigma_Q^{+3/2}}, & {\bf \Sigma_Q^{+1/2}}, &
{\bf \Sigma_Q^{-1/2}}, \\ & {\bf \Lambda_Q^{+1/2}}, &  \nonumber
\end{eqnarray}
defined by \begin{eqnarray}  {\bf [Q u u]_{3/2}}, & {\bf [Q (u
d)_s]_{1/2}}, &  {\bf [Q d d]_{-1/2}}, \\ & {\bf  [Q (u
d)_a]_{1/2}}, & \nonumber
\end{eqnarray}  and an iso-doublet of mesons
\begin{equation} {\bf Q_u^{+1/2} = [\bar Q u]_{+1/2}, \; Q_d^{-1/2} = [\bar Q d]_{-1/2} }.
\end{equation}

Searches for fractionally charged heavy particles exclude them with
mass less than 200 GeV
\cite{Acosta:2002ju,Perl:2004qc,Fairbairn:2006gg}. Nevertheless they
can be produced at the Tevatron or the LHC via Drell-Yan processes.
The fermionic exotica are expected to be lighter than their scalar
partners,  due to soft SUSY breaking contributions to the scalar
masses. Therefore the scalar exotica will decay to their fermionic
partners and a gaugino (either a gluino, chargino or neutralino
depending on the quantum numbers of the (s)exotica).  Moreover,
unless the flavor symmetries of the leptonic exotica are broken via
Yukawa couplings to the MSSM Higgs bosons, they will all be
stable.\footnote{Note, bounds on stable heavy hydrogen are very
severe. For example, Ref.~\cite{smith} (see also
Ref.~\cite{Perl:2001xi}) finds the relative abundance to baryons
less than of order $10^{-28}$. Although these fractionally charged
exotics are stable, they can annihilate. We do not consider the
cosmological evolution of these states in this paper.}

The lightest fractionally charged color singlet state will also be
stable.  However, due to exothermic processes\footnote{Similar
exothermic processes for R-hadrons was considered in Refs.
\cite{Kraan:2004tz,Kraan:2005te,Arvanitaki:2005nq}.} such as
$${\bf  \bar Q_d^{+1/2} +  p} \longrightarrow  {\bf \Sigma_Q^{+3/2} + \pi^0}$$ we
would expect the baryonic exotica to be more abundant than the
mesons. In addition we expect the beta decay processes $$ {\bf
\Sigma_Q^{-1/2}} \longrightarrow {\bf \Lambda_Q^{+1/2} + e^- + \bar
\nu_e} $$ to occur with the lifetime ${\bf \tau_{\Sigma_Q^{-1/2}}}
\sim 10^{-5}$ sec., of order ${\bf \tau_{\Sigma^+}/Br(\Sigma^+
\rightarrow \Lambda^0 + e^+ + \nu_e)}$. Thus the ${\bf Q}$s will be
produced at the Tevatron or the LHC predominantly via gluons, they
will hadronize and the color singlet bound state exotica will likely
be stopped in the hadronic calorimeter.  Note, the electric charge
on one fractionally charged state cannot be screened. However two
fractionally charged states can have their charges screened by the
surrounding normal matter.

The novel feature of these {\em exotica} is that {\em even though
they do not come in complete representations of $\text{SU}(5)$, they
nevertheless preserve gauge coupling unification at one loop.}

\subsection{Gauge coupling unification}

Consider first the evaluation of $\sin^2(\theta_W)$ at $M_{\text{GUT}}$. In
general we have
\begin{eqnarray}  & \sin^2(\theta_W)\big|_{M_G} =  1/ (1 + C^2) & {\rm with} \\
& \displaystyle C^2 =  \frac{\text{Tr} (Y^2/4)}{\text{Tr} (T_{3 L}^2)} & . \label{eq:C}
\end{eqnarray}
For example, if we take the trace over the 5-plet of $\text{SU}(5)$ we find
$C^2 = 5/3$ and $\sin^2(\theta_W)|_{M_G} = 3/8.$   However, we find
the same value of $C^2 = 5/3$ for the exotica in Eqn.
\ref{eq:exotica}.   Hence the GUT boundary conditions are unchanged
with the addition of these states.

Moreover the RG running below the GUT scale is also unchanged. At
one loop we have
$$ \frac{d \alpha_i}{d t} = - \frac{b_i}{2 \pi} \alpha_i^2, \;\; t = \ln
(\mu/\mu_0) $$  with
\begin{eqnarray}  b_i =  3 C_2(G_i) - n T(r).
\label{eq:RGrunning}
\end{eqnarray}  In this equation, $n$ is the number of chiral multiplets in the
representation $r$ of the gauge group $G_i$.  $\text{Tr} (T_A T_B) = T(r)
\delta_{A B}$, where $T_A$ is the gauge generator for the chiral
multiplet.   $T(r) = 1/2$ for the $N$ dimensional representation of
$\text{SU}(N)$ and $T(r) = (3/5) Y^2 /4$ for $\text{U}(1)_Y$.   Finally for
$\text{SU}(N)$, the quadratic Casimir for the adjoint representation,
$C_2(\text{SU}(N)) = N$.

Let $\Delta b_i$ be the contribution of the exotica to the RG
running parameters, Eqn. \ref{eq:RGrunning}.    We find $\Delta b_i
= - 1$ for all $i$.  Hence,  gauge coupling unification is
unaffected by these states at one loop,  even if they have weak
scale masses.

One may inquire, whether these novel exotica are generic in string
model constructions.  We have looked at the exotica found in the
``mini-landscape" search of Ref.~\cite{Lebedev:2006kn}.  In many
cases (of order 5\%) we find exotica with properties similar to
those in Eqn. \ref{eq:exotica}.  However, unlike the previous
exotica, these states transform non-trivially under a hidden sector
gauge group.
 Consider the following two examples:

\bigskip

$\bullet$\\[-5ex]
 \begin{minipage}{\linewidth}\begin{eqnarray}
2\times [(\bsb{3}, \bs{1}, \bs{1})_{-1/3}     &+& (\bs{1}, \bs{1}, \bs{1})_{1}      \label{eq:exotica2} \\
+ (\bs{1}, \bs{2}, \bs{1})_0                  &+& (\bs{1}, \bs{1}, \bs{1})_{\pm 1}] \nonumber \\
+ 2 \times [(\bs{3}, \bs{1}, \bs{1})_{1/3}    &+& (\bs{1}, \bs{1}, \bs{1})_{-1}     \nonumber \\
+ (\bs{1}, \bs{2}, \bs{1})_0 ]                &+& (\bs{1}, \bs{1}, \bs{2})_{\pm 1}  \nonumber
\end{eqnarray} \end{minipage}
transforming under $$\text{SU}(3)_c \times \text{SU}(2)_L \times \text{SU}(2)_{A}
\times \text{U}(1)_Y, $$ and

\bigskip

$\bullet$ \\[-5ex]
\begin{minipage}{\linewidth}\begin{eqnarray}
[(\bs{3}, \bs{2}, \bs{1}, \bs{1})_{1/3}       &+& 5 \times (\bsb{3}, \bs{1}, \bs{1}, \bs{1})_{2/3} \label{eq:exotica3} \\
+ 2 \times (\bs{1}, \bs{2}, \bs{2}, \bs{1})_0 &+& 4 \times (\bs{1}, \bs{1}, \bs{2}, \bs{1})_{\pm 1}] \nonumber \\
+ [(\bsb{3}, \bs{2}, \bs{1}, \bs{1})_{-1/3}   &+& 5 \times (\bs{3}, \bs{1}, \bs{1}, \bs{1})_{-2/3}  \nonumber\\
+ 2 \times (\bs{1}, \bs{2}, \bs{1}, \bs{2})_0 &+& 4 \times (\bs{1}, \bs{1}, \bs{1}, \bs{2})_{\pm1}] \nonumber
\end{eqnarray}\end{minipage}
transforming under $$\text{SU}(3)_c \times \text{SU}(2)_L
\times \text{SU}(2)_{A}\times \text{SU}(2)_{B}\times \text{U}(1)_Y. $$

These states can all be given a supersymmetric mass ${\bf M}$
without breaking any gauge symmetries.  Moreover the low energy
theory is anomaly free.   Although the exotica do not affect the
value of the GUT scale or the low energy prediction from gauge
coupling unification at one loop, they do increase the value of
$\alpha_{GUT}$.  In the first example (Eqn. \ref{eq:exotica}) we
find $\alpha_{GUT}^{-1} = 19$.   In the second example (Eqn.
\ref{eq:exotica2}) we find $\alpha_{GUT}^{-1} = 14$. These two cases
are consistent with perturbative unification.  However, in the third
example (Eqn. \ref{eq:exotica3}), there is a problem since we reach
the Landau pole before the GUT scale.  Hence, if we demand
perturbative unification, this last case is excluded.

In these examples, the hidden $\text{SU}(2)$ gauge symmetry can become
strong forming $\text{SU}(2)$ singlet bound states.   We consider a more
general example below.

\subsection{Exotica with Hidden sector charge}

As discussed in the above examples it is possible in string theory
for the exotica to transform non-trivially under a hidden sector
gauge group.   Here we consider a generalized example, not obtained
from a particular string construction, which has interesting
phenomenology. Consider a hidden sector gauge group $\text{SU}(N)$
with the exotica transforming as
\begin{equation} [(\bs{6}, \bs{1}, \bs{N}) + (\bs{1}, \bs{2}, \bs{N})] + \text{c.c.}  \label{eq:Nexotica2} \end{equation}
under \begin{equation} \text{SU}(6) \times \text{SU}(2)_R \times \text{SU}(N).
\label{eq:Nexotica1}
\end{equation}
Or under \begin{equation} \text{SU}(4) \times \text{SU}(2)_L \times \text{SU}(2)_R \times
\text{SU}(N), \label{eq:NexoticaPS1}
\end{equation} the exotica transform as
\begin{equation} [(\bs{4}, \bs{1}, \bs{1}, \bs{N}) + (\bs{1}, \bs{2}, \bs{1}, \bs{N}) + (\bs{1}, \bs{1}, \bs{2}, \bs{N})] + \text{c.c.}  \label{eq:NexoticaPS2} \end{equation}
Note, pursuant to the previous section, values of $N > 3$ are
excluded by demanding perturbative unification.

Assuming the hidden sector gauge coupling gets strong at a
scale $\Lambda_N \gg M_Z$, the exotica will form $\text{SU}(N)$ singlet
bound states with mass of order $\Lambda_N$. Once again these states
will not affect gauge coupling unification at one loop.   The
phenomenology of such $\text{SU}(N)$ singlet ``baryons" and ``mesons" will
depend on the values of $N$ and $\Lambda_N$.

At the scale $\Lambda_N$ we can expect the N-exotica (Eqn.
\ref{eq:Nexotica2}) and their charge conjugates to form chiral
condensates.\footnote{Unlike technicolor theories,  these chiral
condensates do not break the Standard Model gauge symmetries.}
Neglecting the Standard Model gauge interactions, this theory has a
$\text{U}(8)_L \times \text{U}(8)_R$ chiral symmetry which is broken
to $\text{U}(8)_{\text{vector}}$ via the condensates. Hence we
obtain 64 pseudo-Nambu-Goldstone [PNG] bosons, expected to be the
lightest particles of the strong $\text{SU}(N)$ gauge interactions.
For an analysis of the PNG spectrum, see
\cite{Dimopoulos:1979sp,Dimopoulos:1979za,Dimopoulos:1980yf}. Most
of these states transform non-trivially under QCD and are expected
to have mass  $m \approx (\frac{C_2(r)
\alpha_s(\Lambda_N)}{\alpha_{EM}}) \
\frac{(m_{\pi^+}^2-m_{\pi^0}^2)^{1/2}}{\Lambda_{\text{QCD}}} \
\Lambda_N \sim  \Lambda_N$ where $C_2(r)$ is the quadratic Casimir
for the pseudo-NG boson in representation $r$ of QCD.  On the other
hand, some states with only electroweak quantum numbers will be much
lighter.  These states typically have mass squared of order
\begin{equation} m^2 \approx \alpha_2(\Lambda_N) \Lambda_N^2.  \end{equation}  However, a
few of them (i.e. bound states $[(\bs{1}, \bs{2}, \bs{1}, \bsb{N}) \otimes (\bs{1}, \bs{1}, \bs{2},
\bs{N})]$ under [Pati-Salam$\,\,\times \,\,\text{SU}(N)$] have mass squared
\begin{equation} m^2 \approx \frac{3 \alpha}{4 \pi} M_Z^2
\ln{(\Lambda_N^2/M_Z^2)}. \label{eq:light}
\end{equation} These states have identical electroweak quantum numbers to
the pair of Higgs doublets in the MSSM, i.e.  \v{H}$_u$, \ \v{H}$_d$
and, hopefully without confusion, we use this notation to apply to
these exotica here. Finally the lightest states are axion-like with
no Standard Model gauge quantum numbers.

These light states significantly constrain the lower bound on
$\Lambda_N$. In fact, bounds on an invisible axion would apply here
and we would therefore expect $\Lambda_N > 10^8$ GeV, from
astrophysical constraints on an invisible axion
\cite{Sikivie:2005zz}.  However, the charged states \v{H}$_u^+$
and \v{H}$_d^-$ would have mass less than $M_Z$ even for
$\Lambda_N \sim M_{\text{GUT}}$ (see Eqn. \ref{eq:light}). They would most
likely have been observed at LEP or the Tevatron via the Drell-Yan
production of the exotica through an off-shell photon or $Z$.  The
only way out is to also give these exotica (Eqn. \ref{eq:Nexotica2})
a gauge invariant mass ${\bf M} \geq 200$ GeV.    Then all the
$\text{SU}(N)$ gauge singlets get mass of order ${\bf M}$.

We then can consider two possibilities,  either $\Lambda_N \geq {\bf
M}$ or $\Lambda_N \ll {\bf M}$.   The first case is comparable to
QCD with all quark masses less than or equal to
$\Lambda_{\text{QCD}}$. The PNG bosons will then have two
contributions to their mass squared, i.e. the radiative
contributions considered previously and the explicit mass
contribution given by $\delta m^2 = {\bf M} \Lambda_N^3/F_N^2$ where
$F_N$ is the axion decay constant. A more novel scenario occurs if
we take $\Lambda_N \ll {\bf M}$. These exotica have properties
similar to the ``quirks" introduced in Ref.~\cite{Kang:2006yd}. They
can be produced at the LHC.  When the exotics are produced they can
separate by large distances in the detector before forming the bound
state, since their effective string tension is so much smaller than
their mass. In addition, we expect \v{H}$_u^+$ and \v{H}$_d^-$ to be
heavier than their neutral weak doublet partners \v{H}$_u^0$ and
\v{H}$_d^0$. Thus the charged states will decay via the beta process
\v{H}$_d^- \rightarrow \,\,\,$\v{H}$_d^0 \,\,+ \,\,e \,\,+ \,\, \bar
\nu_e$. The neutral ones are stable and may be dark matter
candidates.\footnote{These neutral states would be comparable to the
$K^0$s in the Standard Model, assuming strangeness was conserved.}
Finally, the lightest $\text{SU}(N)$ singlet exotics can be defined
as follows. Let $\Psi = (8, N)$ define a Dirac spinor for the
exotica of Eqn. \ref{eq:Nexotica2} in an $\text{SU}(8) \times
\text{SU}(N)$ notation. Then the axial currents $J^A_{\mu 5} = \bar
\Psi \gamma_\mu \gamma_5 T^A \Psi$, with $T^A$ given by the 4
generators in the Cartan sub-algebra of $\text{SU}(8)$ but not in
$\text{SU}(3)_c \times \text{SU}(2)_L$ , create these light PNG
bosons from the vacuum. Two of these can either decay to two photons
or two gluons through the triangle anomaly. Clearly, all these
exotica would have very interesting signatures at the LHC.

\section{Conclusion}

It is well known that the prediction of gauge coupling unification
in traditional SUSY GUTs is unaffected at one loop by the presence
of light vector-like states in complete multiplets of
$\text{SU}(5)$. In this letter we have identified a class of
vector-like {\em exotica} which\\[-4ex]
\begin{itemize}
\item do not come in complete $\text{SU}(5)$ multiplets,\\[-4ex]
\item do not affect gauge coupling unification at one loop,\\[-4ex]
\item  can have fractional electric charge, and\\[-4ex]
\item are found in orbifold constructions of the heterotic string.\\[-4ex]
\end{itemize}
Such states provide an interesting challenge for the LHC.

\textbf{Acknowledgments.} We would like to acknowledge research
supported in part by the Department of Energy under Grant No.\
DOE/ER/01545-872.

\providecommand{\bysame}{\leavevmode\hbox to3em{\hrulefill}\thinspace}

\end{document}